# A technical note for a Shor's algorithm by phase estimation

G. Fleury and P. Lacomme

*Université Clermont-Auvergne, Clermont-Auvergne-INP,*
*LIMOS (UMR CNRS 6158),*
*1 rue de la Chebarde,*
*63178 Aubière Cedex, France*

*gerard.fleury@isima.fr, philippe.lacomme@isima.fr*

## 1. Introduction

The most famous quantum algorithm is the Shor's algorithms that is dedicated to integer factorization and that received a considerable amount of attention due to its possible application in the cryptanalysis field. The algorithm proposes a super-polynomial execution speedup as regards the classical resolution on classical computers. The RSA encryption method lies on the hypothesis that it is exponentially harder to factor large numbers that achieving a multiplication ensuring secure send of information online.

Feynman has been the very first researcher considering that quantum mechanics could be more efficient computationally than a Turing machine considering that a computer based on quantum mechanics should avoid the expensive computation required for simulation on a classical computer. The question has been addressed first by (Deutsch and Jozsa, 1992) proving that some problems can be solved fast to optimality using quantum computers. In 1994 (Simon, 1994) introduced a circuit using an oracle to solve a problem (i.e. a black-box that can be checked) requiring a polynomial time since it requires an exponential time on a classical non quantum computer.

(Shor, 1994) introduced a quantum computer algorithm for factoring a integer number $N$ undirectly by computing the order of an element $a$ in the multiplicative group ($mod\ N$) considering the lower integer $l\ /\ a^l \equiv 1\ mod\ N$. He defines a new promising approach to factor a number into a product of primes and his proposal resulted in a significant improvement of the factoring algorithms efficiency. The corner stone of the Shor's algorithm is the modular exponentiation that is the most computational component (in time and space) of Shor's algorithm.

## 2. Proposition

Finding a factor of $N$ ($N$ is an odd number) requires a method to compute the order $l$ of $a$ where $a$ is a random integer number / $a \in [2; N-1]$:

$$a^l \equiv 1\ mod\ N$$
$$a^l - 1 \equiv 0\ mod\ N$$

If $l$ is an even number we have:

$$\left(a^{l/2} - 1\right).\left(a^{l/2} + 1\right) \equiv 0\ mod\ N$$

and the Euclidean algorithm of (Knuth, 1981) can be used to compute the $gcd(a^{\frac{l}{2}} - 1; N)$ and $gcd(a^{\frac{l}{2}} + 1; N)$ giving the two dividers of $N$. If $l$ is an odd number, it is possible to investigate both the even numbers $l - 1$ and $l + 1$.

Our work propose an architecture where the computation of the order of an element $a$ in the multiplicative group ($mod\ N$) is replaced by the computation of the order of $a$ in the phase additive group ($mod\ 2\pi$). The Quantum Circuit defines a probability phase distribution that aggregates probabilities on phases that indirectly model order of $a$. The measurement gates jointly gives a float value that defines the phase that will permit to find the order of a with a high probability through further classical post-processing. The sampling of the probability distribution permits to create a heap $\chi$ of set of phases (step 5 to 8 in algorithm 1). The post-processing consists in iteratively considering the phases registered in $\chi$ which may gives on order of $a$: $l_i = \chi_i \times N$.

If $l_i \equiv 0\ mod\ 2$ the post-processing consists in computing both $gcd(a^{l_i/2} - 1; N)$ and $gcd(a^{l_i/2} + 1; N)$ which define two dividers of $N$ and that are saved into the set of divider of $N$ referred to as $\wp$. If $l_i$ is an odd number, both the even number $l_i - 1$ and the even number $l_i + 1$ are investigated for giving factors of $N$.

Note that the "necessary" conditions at steps 8, 9 and 24 have to be design over considerations of the objective function that could be, but not limited to: 1) computation of one divider only; 2) computation of a fixed number of dividers; 3) computation of all dividers.

**Algorithm 1. Shor's algorithm based on phase**

```
Input parameter
      N : a positive integer number
      ns : maximal number of measurement (sampling)
Output parameter:
      χ : set of phases
      ℘ : set of dividers of N
begin
   1. χ = ∅ ; ℘ = ∅
   2. repeat
   3.    a := random (2; N − 1)
   4.    |φ_l⟩ := Shor_quantum_circuit (a, N)
   5.    repeat
   6.       phase φ_l = Measurement of the circuit
   7.       χ = χ ∪ φ_l
   8.    until (necessary) or (iterations exceed ns)
   9.    while (i in 0 to card(χ)) and (necessary) do
  10.       l_i = χ_i × N
  11.       if l_i ≡ 0 mod 2 then
  12.          Compute v_1 = a^(l/2) + 1 and v_2 = a^(l/2) − 1
  13.          p_1 = gcd(v_1; N) ; p_2 = gcd(v_2; N)
  14.          ℘ = ℘ ∪ {p_1, p_2}
  15.       end if
  16.       if l_i ≡ 1 mod 2 then
  17.          Compute v_1 = a^((l−1)/2) + 1 and v_2 = a^((l−1)/2) − 1
  18.          Compute v_3 = a^((l+1)/2) + 1 and v_4 = a^((l+1)/2) − 1
  19.          p_1 = gcd(v_1; N) ; p_2 = gcd(v_2; N)
  20.          p_3 = gcd(v_3; N) ; p_4 = gcd(v_4; N)
  21.          ℘ = ℘ ∪ {p_1, p_2, p_3, p_4}
  22.       end if
  23    end while
  24 until (necessary)
end
```

## 2.1. Shor's Quantum Circuit based on phase

Given $a$, to find $l$ such that $a^l \equiv 1 \bmod N$, the circuit is based on the following considerations. First we find $p$ the lowest power of 2 such that $2^p = q$ satisfy $2^p > N + 1$ and we compute $n$ the lowest power of 2 such that $2^n > a$.

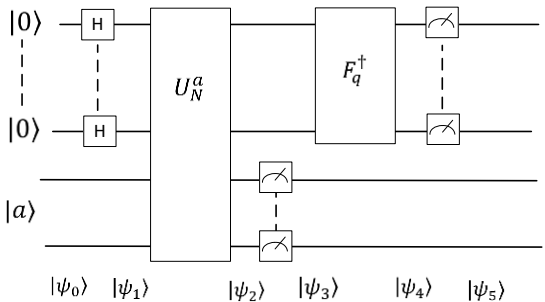

Fig. 1. Quantum Circuit

The second step is easy since its entails is putting each qubit of the first register into the superposition $\frac{1}{\sqrt{2}}(|0⟩ + |1⟩)$ and to compute $a^l \bmod N$ in the second register leaving the circuit in the state:

$$|\psi⟩ = \frac{1}{\sqrt{q}} \cdot \sum_{j=0}^{q-1} |j⟩ \otimes |a⟩$$

The most computational quantum part of the classical Shor's algorithm is the modular exponentiation $U_N^a$ that achieves the transformation to its quantum input:

$$U_N^a \cdot |l⟩ \otimes |y⟩ = |l⟩ \otimes |y \oplus (a^l \bmod N)⟩$$

with $\quad 0 \leq l < q$
$\quad\quad 0 \leq y < 2^n$

The Hadamard gates on the first $q$ qubits creates a superposition of number $l$ before application of the modular exponentiation.

$$|\psi_2⟩ = U_N^a \cdot \left( \frac{1}{\sqrt{q}} \sum_{j=0}^{q-1} |j⟩ \otimes |0⟩^{\otimes n} \right)$$

$$|\psi_2⟩ = U_N^a \cdot \left( \frac{1}{\sqrt{q}} \sum_{j=0}^{q-1} |j⟩ \otimes |0 \ldots 0⟩ \right)$$

$$|\psi_2\rangle = \frac{1}{\sqrt{q}} \sum_{j=0}^{q-1} |j\rangle \otimes |\vec{0} \oplus a^l \bmod N\rangle$$

Classical modular exponentiations are based on: 1) controlled multiplier/accumulator (Draper et al., 1998) (Bauregard et al., 2003) 2) quantum divider based on the algorithm introduced by (Grandlund and Montgomery, 1994).

We propose to replace the modular exponentiation $U_N^a$ by $U_N^{\varphi_a}$ that achieves the transformation to its quantum input in:

$$U_N^{\varphi_a} . |l\rangle \otimes |y\rangle = |l\rangle \otimes |y \oplus \varphi_{a^l} \bmod 2\pi)\rangle$$

with $\quad 0 \leq l < q$
$\quad\quad 0 \leq y < 2^n$

The Hadamard gates at the first $q$ qubits have created a superposition of number $l$ before application of the modular exponentiation.

$$|\psi_2\rangle = U_N^{\varphi_a} . \left( \frac{1}{\sqrt{q}} \sum_{j=0}^{q-1} |j\rangle \otimes |0\rangle^{\otimes n} \right)$$

$$|\psi_2\rangle = U_N^{\varphi_a} . \left( \frac{1}{\sqrt{q}} \sum_{j=0}^{q-1} |j\rangle \otimes |0 \dots 0\rangle \right)$$

$$|\psi_2\rangle = \frac{1}{\sqrt{q}} \sum_{j=0}^{q-1} |j\rangle \otimes |\vec{0} \oplus \varphi_{a^l} \bmod 2\pi\rangle$$

The classical post-processing of the measurement in the computational basis after the Inverse Quantum Fourier Transform ($QFT^{-1}$ block) is a sampling procedure that gives with high probability the period of the function $f(l) = \varphi_a^l \bmod 2\pi$.

During post-processing, all the phases $\varphi_a$ are saved into a pool $\chi$ of phases and all phases are iteratively investigated to check if the phase can permit to obtain one divider:

For all $l_i = \chi_i \times N$ / $l_i \equiv 0 \bmod 2$, compute $v_1 = a^{l_i/2} - 1$ and $v_2 = a^{l_i/2} + 1$ and next $p_1 = gcd(v_1; N)$ and $p_2 = gcd(v_1; N)$. $p_1$ and $p_2$ are both dividers of $N$.

For all $l_i = \chi_i \times N$ / $l_i \equiv 1 \bmod 2$, compute $v_1 = a^{(l_i+1)/2} - 1$, $v_2 = a^{(l_i+1)/2} + 1$, $v_3 = a^{(l_i-1)/2} - 1$, $v_4 = a^{(l_i-1)/2} + 1$. Compute next $p_1 = gcd(v_1; N)$, $p_2 = gcd(v_2; N)$, $p_3 = gcd(v_3; N)$ and $p_4 = gcd(v_4; N)$ that are dividers of $N$.

2.2. Quantum Fourier Transformation

For $p$ qubits, $q = 2^p$ defines $2^p$ basis vectors and the Quantum Fourier Transformation is:

$$F_q(|k\rangle) = X_{|k\rangle} = \frac{1}{\sqrt{2^n}} \sum_{j=0}^{2^n-1} e^{2.\pi.i.\frac{k.j}{2^n}} . |l\rangle$$

$$F_q(|k\rangle) = X_{|k\rangle} = \frac{1}{\sqrt{q}} \sum_{j=0}^{q-1} e^{2.\pi.i.\frac{k.j}{2^n}} . |l\rangle$$

$$F_q(|k\rangle) = X_{|k\rangle} = \frac{1}{\sqrt{q}} \sum_{j=0}^{q-1} \omega^{k.j} . |j\rangle \text{ avec } \omega^{k.j} = e^{2.\pi.i.\frac{k.j}{2^n}}$$

By consequence

$$F_q^\dagger(|k\rangle) = \frac{1}{\sqrt{q}} \sum_{j=0}^{q-1} \omega^{-k.j} . |j\rangle$$

2.3. *Modular exponentiation of the phase : $U_N^{\varphi_a}$*

The design uses $q = 2^p$ qubits to represents l, a controlled qubit for the Inverse Fourier Transformation, $n$ qubits to represents $a$ and a controlled qubit required by $U_N^{\varphi_a}$ (figure 2).

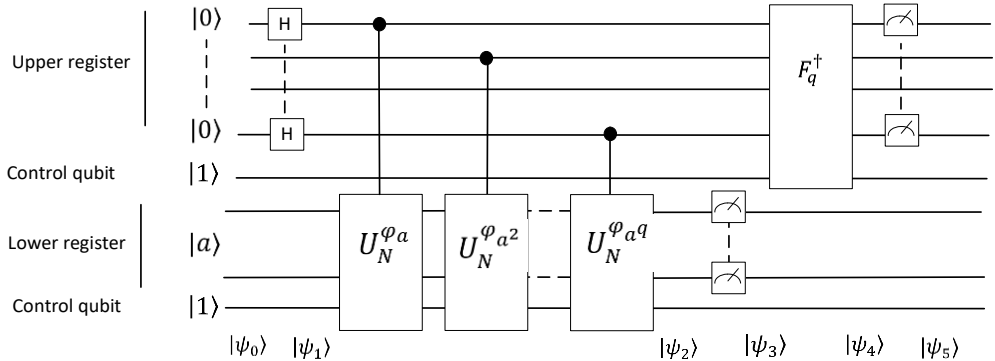

Fig. 2. Design of the exponentiation circuit using controlled modular phase

The design uses $q$ controlled-$U$ ($CU$) block, and each block referred to as $U_N^{\varphi_{a^{2^j}}}$ is a controlled modular $2.\pi$ multiplier of its input (the lower quantum register) by constant $\varphi_a$ by $(2^{j-1} \bmod N)$. Each block is controlled by the corresponding $|l_j\rangle$ qubit of the upper quantum register.

Definition of $\varphi_{a^{2^j}}$

Let us note $\varphi_a = \frac{2.\pi}{N}.a$

The phase of $a$ is:
$$\varphi_{a^{2^j}} = (2^{j-1} \bmod N).\frac{2.\pi}{N}.a$$

The modular exponentiation $U_N^{\varphi_{a^k}}$ requires the iterative computation of
$$\varphi_{a^1}, \varphi_{a^2}, \varphi_{a^4}, \varphi_{a^8}, \ldots, \varphi_{a^k}, \ldots$$
*i.e.* the computation of
$$\varphi_{a^{2^0}}, \varphi_{a^{2^1}}, \varphi_{a^{2^2}}, \varphi_{a^{2^3}}, \ldots, \varphi_{a^{2^{p-1}}}, \varphi_{a^{2^p}}, \ldots$$
We note
$$\varphi_{a^{2^p}} = \varphi_{a^{2^{p-1}}} + \sum_{k=1}^{2^{p-1}} \varphi_a = \varphi_{a^{2^{p-1}}} + 2^{p-1}.\varphi_a$$

For example with $a = 3$, we have $\varphi_a = \frac{2.\pi}{N}.3$, $\varphi_{a^2} = \frac{2.\pi}{N}.3 + \frac{2.\pi}{N}.3$ and
$$\varphi_{a^4} = \frac{2.\pi}{N}.3 + \frac{2.\pi}{N}.3 + \frac{2.\pi}{N}.3 + \frac{2.\pi}{N}.3$$
$$\varphi_{a^4} = \varphi_{a^2} + 2^{2-1}.\varphi_a = \varphi_{a^2} + 2.\varphi_a$$
We have next:
$$\varphi_{a^8} = \varphi_{a^4} + 2^2.\varphi_a = \varphi_{a^4} + 4.\varphi_a = 8.\varphi_a$$
$$\varphi_{a^{16}} = \varphi_{a^8} + 2^3.\varphi_a = \varphi_{a^8} + 8.\varphi_a = 16.\varphi_a$$

Definition of $U_N^{\varphi_{a^k}}$

A controlled $U_N^{\varphi_{a^k}}$ block is based on the application of $CP(\alpha)$ conditional phase rotation gates that need to be repeated a logarithmic number of times regarding to $N$ and next on the application of the Inverse Fourier Transformation (Fig. 3). The diagonal symmetric controlled gate $CP(\alpha)$ is used to induce a phase on the state of the target qubit number $n$ depending on the control qubit state.

Matrix representation of $CP(\alpha)$ for two qubits
$$CP(\alpha) = |0\rangle\langle 0| \otimes I + |1\rangle\langle 1| \otimes P = \begin{pmatrix} 1 & & & \\ & 1 & & \\ & & 1 & \\ & & & e^\alpha \end{pmatrix}$$

$n$ consecutive applications of $CP(\alpha)$ gates are required for defining a $n$-qubit quantum number
$$|b\rangle = |b_{n-1}\rangle|b_{n-2}\rangle \ldots |b_1\rangle|b_0\rangle$$
that is transformed by a simple uncontrolled Inverse Quantum Fourier Transformation into a quantum state
$$|\varphi_{a^k}\rangle = |\varphi_{a^k}^{n-1}\rangle|\varphi_{a^k}^{n-2}\rangle|\varphi_{a^k}^{n-3}\rangle \ldots |\varphi_{a^k}^0\rangle$$

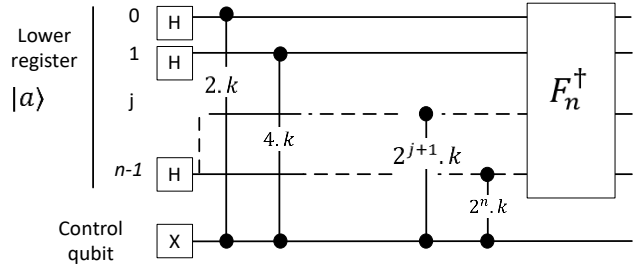

Fig. 3. Design of $U_N^{\varphi_{a^k}}$

We have:
$$F_n(|b\rangle) = \frac{1}{\sqrt{2^n}} \sum_{j=0}^{2^n-1} e^{2.\pi.i.\frac{b.j}{2^n}}.|j\rangle$$

The individual $j^{th}$ qubit $|\varphi_{a^k}^j\rangle$ of $|\varphi_{a^k}\rangle$:
$$|\varphi_{a^k}^j\rangle = \frac{1}{2}(|0\rangle + e^{2.\pi.i.\frac{b}{2^j}}|1\rangle)$$

## 3. Experimental validation

With the following examples, the goal is to illustrate the performances of this phase based Shor's algorithm on both simulator and physical quantum optimizer. Experiments are achieved considering 5 large numbers (introduced in table 1) with a number of dividers that varies from 4 to 24.

Table 1. Integer number used for experimentations

| N | # of dividers | Dividers |
|---|---|---|
| 844 821 | 24 | 1; 3; 9; 37; 43; 59; 111; 129; 177; 333; 387; 531; 1591; 2183; 2537; 4773; 6549; 7611; 14319; 19647; 22833; 93869; 281607; 844821 |
| 1 414 583 | 4 | 1; 821; 1723; 1414583 |
| 1 660 759 | 4 | 1; 1129; 1471; 1660759 |
| 3 131 759 | 4 | 1; 1471; 2129; 3131759 |
| 5 131 763 | 8 | 1; 7; 13; 91; 56393; 394751; 733109; 5131763 |
| 5 131 769 | 4 | 1; 103; 49823; 5131769 |
| 7 131 763 | 4 | 1; 223; 31981; 7131763 |
| 7 131 769 | 4 | 1; 113; 63113; 7131769 |
| 17 131 761 | 6 | 1; 3; 9; 1903529; 5710587; 17131761 |
| 327 131 761 | 16 | 1; 11; 73; 263; 803; 1549; 2893; 17039; 19199; 113077; 211189; 407387; 1243847; 4481257; 29739251; 327131761 |
| 327 131 767 | 8 | 1; 229; 233; 6131; 53357; 1403999; 1428523; 327131767 |
| 1 593 389 361 | 8 | 1; 3; 21121; 25147; 63363; 75441; 531129787; 1593389361 |
| 1 593 389 363 | 4 | 1; 31981; 49823; 1593389363 |
| 298 500 156 599 | 4 | 1; 139; 930049; 320951; 320951; 298500156599 |

The number of qubits required depends on both $N$ and $a$ : the experiments on the simulator are limited to 32 qubits and the experiments on quantum optimizer are limited to 124 qubits that is the largest IBM physical quantum optimizer available. A number with many dividers should be easier to factor in the sense that one circuit execution should permit to obtain several dividers in one sampling run. Note that for a theoretical point of view, the value of $a$ should be defined at random but from a practical point of view, an efficient implementation should investigate small values first to obtain smaller circuits.

Experiments have been achieved on the Brooklyn Quantum Physical Optimizers introduced in table 2 and has been chosen taking into account the individual optimizers availabilities and the total pending jobs involved in experiments with the potential to lead the results in acceptable delays at the particular moment. Note that preliminary experiments achieved on Auckland, Montreal and Washington quantum optimizers push us into considering that very similar results should be obtained. Let us note that the Brooklyn physical quantum optimizer is not the best one in term of number of qubits and it is not the more efficient as regards the quantum volume.

Table 2. Examples of physical quantum optimizers

| Physical quantum optimizer | # of qubits | QV |
|---|---|---|
| auckland | 27 | 64 |
| montreal | 27 | 128 |
| brooklyn | 65 | 32 |
| washington | 127 | 64 |

3.1. Numerical experiments with 1591

For the small $N = 1591$ the probability distribution of $l$ are plotted on figure 4. The values $l = 37$, $l = 43$ could occurred when factoring 1591 if $a$ were chosen to be 2 for example. The graphic concerns small $N$ and small $a$ values to make the plot distinguishable but very similar structures of probabilities can be observed for every $N$ and $a$ values. Note that the probabilities plotted on figure 5 are related to an execution on the Brooklyn Physical Quantum Optimizer defining values very close to the values obtained with the simulator.

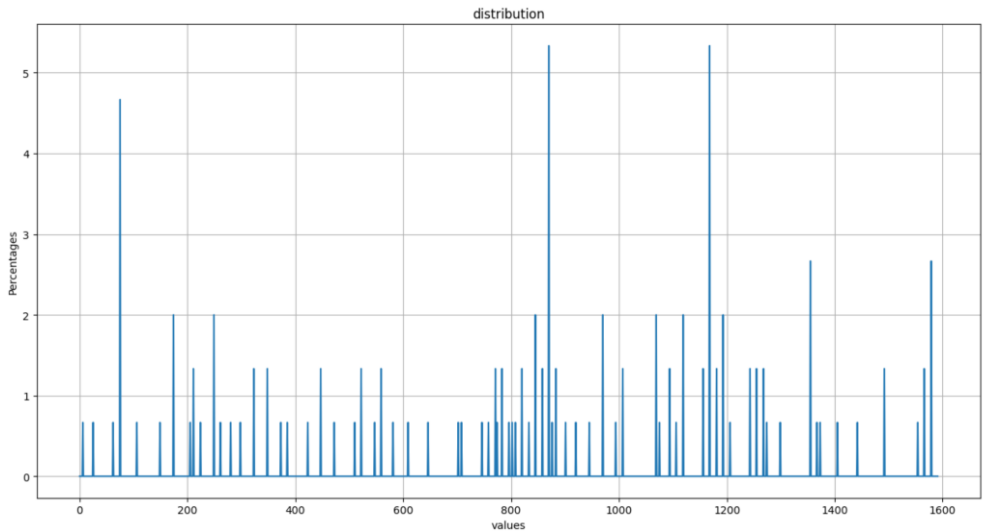

Fig. 4. The probability $P$ to observe values giving $N = 1591$ and $a = 2$ (execution on simulator)

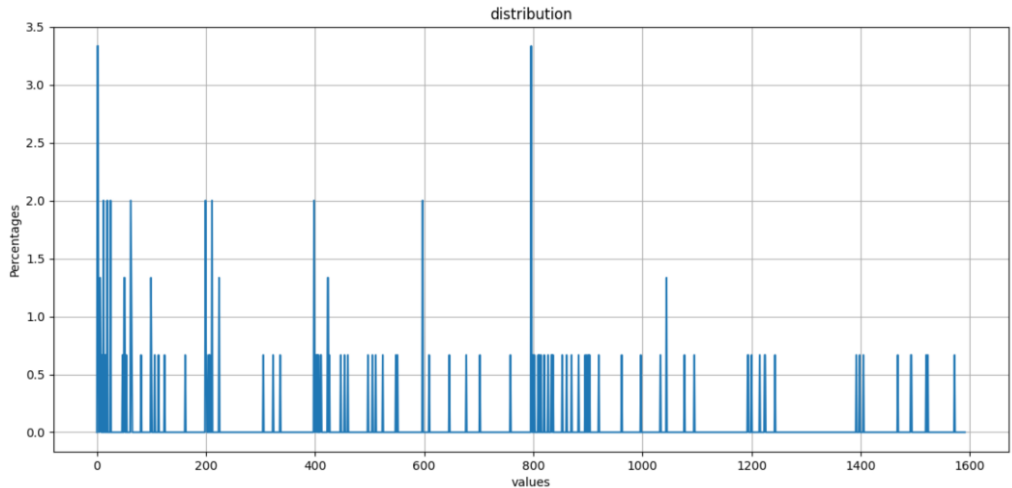

Fig. 5. The probability $P$ to observe values giving $N = 1591$ and $a = 2$ (execution on Brooklyn Physical Quantum Optimizer)

Table 3. Dividers found when factoring 1591 with $a = 2$ (execution on the simulator)

| phase | $l$ | Divider D1 | Divider D1 | $\frac{N}{D1}$ | $\frac{N}{D2}$ |
|---|---|---|---|---|---|
| 0.03515625 | 56 | 43 | 1 | 37 | 1591 |
| 0.16801170509565028 | 267 | 1 | 43 | 1591 | 37 |
| 0.4531679550956503 | 721 | 37 | 1 | 43 | 1591 |
| 0.3788632949043497 | 603 | 1 | 43 | 1591 | 37 |
| 0.6249570449043498 | 994 | 1 | 43 | 1591 | 37 |
| 0.0703125 | 112 | 43 | 1 | 37 | 1591 |
| 0.5898007949043498 | 938 | 1 | 43 | 1591 | 37 |
| 0.9336367050956502 | 1485 | 43 | 1 | 37 | 1591 |
| 0.175824220509565028 | 280 | 43 | 1 | 37 | 1591 |
| 0.16792579490434972 | 267 | 1 | 43 | 1591 | 37 |
| 0.6601132949043498 | 1050 | 1 | 43 | 1591 | 37 |
| 0.4296875 | 684 | 1 | 37 | 1591 | 43 |
| 0.4843320449043497 | 771 | 1 | 43 | 1591 | 37 |
| 0.2265625 | 360 | 37 | 1 | 43 | 1591 |
| 0.3789492050956503 | 603 | 1 | 43 | 1591 | 37 |

The phases in $\chi$ define possible values of $l$ for $a^l$ such that $l$ should be the order of $a$. Table 3 gives the observed value of both the phase and of $l$ obtained when factoring 1591 with $a = 2$: the post-processing enables us to obtain the two non-trivial dividers of 1541 namely 43 and 37.

Execution on the Brooklyn gives quite different probabilities but the observed phase value permits to obtain the two dividers 43 and 37.

3.2. Numerical experiments on simulator when factoring 3 131 759

Execution of the simulator for $N = 3\ 131\ 759$ has been achieved for several values of $a$ and permits to retrieve several dividers depending on the run. With $a = 2$ the runs are unsuccessful and the run 3 with $a = 3$ succeeds retrieving 1471 and 2129 (table 4). Note that the number of measurements (150 here) should be increased to obtain more consistent results between runs.

Table 4. Shor's algorithm on IBM's Simulator for $N = 3\ 131\ 759$

| N | Quantum optimizer | Sampling value | Run | $a$ | Phases | First Divider $D1$ | Second Divider $D2$ | $\frac{N}{D1}$ | $\frac{N}{D2}$ |
|---|---|---|---|---|---|---|---|---|---|
| 3 131 759 | Simulator | 150 | 3 | 3 | 0.9023457562799555 | 1471 | 1 | 2129 | 3131759 |
| | | | 2 | 4 | 0.3945292437200445 | 1 | 2129 | 3131759 | 1471 |
| | | | 3 | 4 | 0.007814506279955507 | 2129 | 1 | 1471 | 3131759 |
| | | | 1 | 5 | 0.3945292437200445 | 1 | 2129 | 3131759 | 1471 |
| | | | 2 | 5 | 0.7988261187200445 | 1 | 2129 | 3131759 | 1471 |
| | | | 1 | 6 | 0.5214863812799555 | 1 | 1471 | 3131759 | 2129 |
| | | | 2 | 6 | 0.5214863812799555 | 1 | 1471 | 3131759 | 2129 |
| | | | 3 | 6 | 0.5214863812799555 | 1 | 1471 | 3131759 | 2129 |
| | | | | 6 | 0.404296875 | 1 | 2129 | 3131759 | 1471 |

*3.3. Numerical experiments for 3 numbers: analysis of the phases and dividers obtained for 3 small numbers*

Table 2 gives the observed values of the phase and the dividers obtained when factoring 844 821 with $a = 2$ during a single run. The results are compliant with the results provided by the simulator and the algorithm succeeds in computing a large part of the dividers.

Experiments for 1 414 583 have been achieved with one run for value of $a$ varying from 2 to 5 and the two dividers are identified with $a = 5$.

The last number 1660759 has two non-trivial dividers only (1 471 and 1 129) that are found for $a = 3, 4, 5$ by the algorithm in one single run as stressed in table 5.

Table 5. Shor's algorithm on IBM's physical quantum optimizer (one run for $a \in [2;5]$): example of results

| N | Physical quantum optimizer | Sampling value | Run | a | Phases | First Divider $D1$ | Second Divider $D2$ | $\frac{N}{D1}$ | $\frac{N}{D2}$ |
|---|---|---|---|---|---|---|---|---|---|
| 844821 | brooklyn | 150 | 1 | 2 | 0.00390625 | 9 | 1 | 93869 | 844821 |
| | | | | | 0.0634765625 | 1 | 3 | 844821 | 281607 |
| | | | | | ... | | | | |
| | | | | | 0.09765625 | 1 | 129 | 844821 | 6549 |
| | | | | | ... | | | | |
| | | | | | 0.0004808439522680786 | 1 | 7611 | 844821 | 111 |
| | | | | | ... | | | | |
| | | | | | 0.1259765625 | 387 | 1 | 2183 | 844821 |
| | | | | | 0.14843006270226808 | 177 | 1 | 4773 | 844821 |
| | | | | | ... | | | | |
| | | | | | 0.0007249845772680786 | 9 | 37 | 93869 | 22833 |
| | | | | | 0.283203125 | 333 | 1 | 2537 | 844821 |
| | | | | | ... | | | | |
| | | | | 3 | 0.06249256270226808 | 1 | 1591 | 844821 | 531 |
| | | | | | 0.0039213166922631715 | 37 | 1 | 22833 | 844821 |
| | | | | | 0.0039272308349609375 | 1 | 43 | 844821 | 19647 |
| | | | | | ... | | | | |
| | | | | | 0.021556854248046875 | 59 | 1 | 14319 | 844821 |
| | | | | | ... | | | | |
| | | | | | 0.06250743729773192 | 129 | 1 | 6549 | 844821 |
| | | | | | 0.06249256270226808 | 387 | 37 | 2183 | 22833 |
| 1414583 | brooklyn | 150 | 1 | 5 | 0.7558975219726562 | 821 | 1 | 1723 | 1414583 |
| 1660759 | brooklyn | 150 | 1 | 3 | 0.5012550354003906 | 1129 | 1 | 1471 | 1660759 |
| | | | | 3 | 0.59498596191406 25 | 1129 | 1 | 1471 | 1660759 |
| | | | | 3 | 0.0003399508056640625 | 1 | 1129 | 1660759 | 1471 |
| | | | 1 | 4 | 0.6427001953125 | 1 | 1129 | 1660759 | 1471 |
| | | | 1 | 5 | 0.0626573876379268 | 1 | 1129 | 1660759 | 1471 |

*3.4. Numerical experiments on IBM programmable noisy quantum optimizer for large number*

An attempt for 3 131 759 factorization is achieved with 5 run and $a$=2 and it succeeds leading to non-trivial dividers i.e. 1471 and 2129 (table 6)

Note that 5 131 763 can be factorized on the Brooklyn Physical Quantum Optimizer and that a single run with $a = 2$ gives several dividers including 1, 7, 13, 91, 56393 , 733109, 394751 and 5 131 763. The algorithm succeeds in factoring 5 131 769 with $a = 2$.

Computation of the dividers of both 7 131 763 and 7 131 769 succeed for small value of $a$ as stressed in table 6.

The algorithm succeeds in factoring the large number 327 131 761 and using several run (5 runs are used in the experiments) several dividers are computed including large dividers. Note that the sampling parameters could have to be tune considering a balance between runs and number of sampling efforts.

Half of the dividers of 327 131 767 are computing on a single run for $a = 2$ including the non-trivial dividers 229, 233, 1 403 999 and 1 428 523.

For number greater that $1.10^9$, the algorithm succeeds as stressed on table 6 where the two dividers of 1 593 389 363 are obtained for $a = 4$ with a single run.

Table 6. Shor's algorithm on IBM's physical quantum optimizer: large numbers

| Number | Physical quantum optimizer | Sampling value | Run | $a$ | Phase | First Divider D1 | Second Divider D2 | $\frac{N}{D1}$ | $\frac{N}{D2}$ |
|---|---|---|---|---|---|---|---|---|---|
| 3 131 759 | brooklyn | 150 | 1 | 2 | 0.445446590172777777 | 1471 | 1 | 2129 | 3131759 |
|  |  |  | 5 | 2 | 0.003911873114575743 | 1471 | 1 | 2129 | 3131759 |
| 5 131 763 | brooklyn | 150 | 1 | 2 | 3.17419498091911e-05 | 7 | 1 | 733109 | 5131763 |
|  |  |  |  | 2 | 2.9293206440808898e-05 | 7 | 1 | 733109 | 5131763 |
|  |  |  |  | 2 | 0.0078125 | 7 | 13 | 733109 | 394751 |
|  |  |  |  |  | ... |  |  |  |  |
|  |  |  |  | 2 | 0.5001835823059082 | 91 | 1 | 56393 | 5131763 |
|  |  |  |  |  | ... |  |  |  |  |
| 5 131 769 | brooklyn | 150 | 1 | 2 | 0.12597545733903687 | 103 | 1 | 49823 | 5131769 |
|  |  |  | 1 | 2 | 0.010017633438110352 | 103 | 1 | 49823 | 5131769 |
|  |  |  |  |  | ... |  |  |  |  |
| 7 131 763 | brooklyn | 150 | 1 | 2 | 0.5019550323486328 | 223 | 1 | 31981 | 7131763 |
|  |  |  |  | 2 | 0.50445556640625 | 223 | 1 | 31981 | 7131763 |
|  |  |  |  |  | ... |  |  |  |  |
| 7 131 769 | brooklyn | 150 | 1 | 2 | 6.2831845453753 | 1 | 113 | 7131769 | 63113 |
|  |  |  |  | 2 | 0.12499911898642438 | 113 | 1 | 63113 | 7131769 |
|  |  |  |  |  | ... |  |  |  |  |
| 17 131 761 | brooklyn | 150 | 1 | 2 | 0.00012243706903525517 | 1 | 3 | 17131761 | 5710587 |
|  |  |  |  | 2 | 0.00012170355596474483 | 3 | 1 | 5710587 | 17131761 |
|  |  |  |  | 2 | 0.5000158937664992 | 9 | 1 | 1903529 | 17131761 |
|  |  |  |  | 2 | 0.5000151602534287 | 9 | 1 | 1903529 | 17131761 |
|  |  |  |  |  | ... |  |  |  |  |
| 327 131 761 | brooklyn | 150 | 1 | 2 | 0.001953650846375354 | 11 | 1 | 29739251 | 327131761 |
|  |  |  | 1 | 2 | 0.0019550515555275755 | 73 | 1 | 4481257 | 327131761 |
|  |  |  | 1 | 2 | 0.03128812776576149 | 803 | 1 | 407387 | 327131761 |
|  |  |  | 1 | 2 | 0.03137252794276344 | 1 | 11 | 327131761 | 29739251 |
|  |  |  |  |  | ... |  |  |  |  |
|  |  |  | 3 | 2 | 0.12744569778442383 | 263 | 1 | 1243847 | 327131761 |
|  |  |  |  |  | ... |  |  |  |  |
|  |  |  | 4 | 2 | 0.6250076120502817 | 1243847 | 1 | 263 | 327131761 |
|  |  |  |  |  | ... |  |  |  |  |
|  |  |  | 4 | 2 | 0.7509803964041604 | 2893 | 1 | 113077 | 327131761 |
|  |  |  |  |  | ... |  |  |  |  |
| 327 131 767 | brooklyn | 150 | 1 | 2 | 0.001975993976699339 | 233 | 1 | 1403999 | 327131767 |
|  |  |  | 1 | 2 | 0.0367431640625 | 229 | 1 | 1428523 | 327131767 |
|  |  |  | 1 | 2 | 0.6279297402345071 | 1 | 229 | 327131767 | 1428523 |
|  |  |  |  |  | ... |  |  |  |  |
| 1 593 389 361 | brooklyn | 150 | 1 | 2 | 0.0078125 | 1 | 3 | 1593389361 | 531129787 |
|  |  |  | 1 | 2 | 0.12503055178939176 | 1 | 3 | 1593389361 | 531129787 |
|  |  |  |  |  | ... |  |  |  |  |
| 1 593 389 363 | brooklyn | 150 | 1 | 4 | 0.531250071929831 | 1 | 31981 | 1593389363 | 49823 |
|  |  |  |  |  | ... |  |  |  |  |

## 3. Extra experiments with very large numbers

Experiments have been carried out with the following numbers 298500156599, 237504336099404000, 237504336099404013, and 237504336099404017 that require more than 60 qubits.

To conclude, a set of 3 experiments have been achieved on the Washington physical quantum optimizer using
$$N = 15785823750433609940013$$
$$N = 237515785823750433609940401323457$$
$$N = 2375157858237504336099404013234571$$
All the results are introduced in table 7 and confirm the approach efficiency.

Table 7. Shor's algorithm on IBM's physical quantum optimizer: very large numbers (>60 qubits used): 150 samplings

| N | Run | a | Phases | First Divider $D1$ | Second Divider $D2$ | $\frac{N}{D1}$ | $\frac{N}{D2}$ |
|---|---|---|---|---|---|---|---|
| 2375043360994040000 Brooklyn optimizer 63 qubits | 1 | 2 | 0.25000000838190406 | 275 | 1 | 863652131270560 | 237504336099404000 |
| | | 2 | 2.384190338489134e-07 | 1 | 5 | 237504336099404000 | 47500867219880800 |
| | | 2 | 2.384190338892462e-07 | 1 | 5 | 237504336099404000 | 47500867219880800 |
| | | 2 | 2.3841903383633616e-07 | 5 | 1 | 47500867219880800 | 237504336099404000 |
| | | 2 | 0.25000047869980335 | 5 | 1 | 47500867219880800 | 237504336099404000 |
| | | 2 | 0.25000190832361124 | 5 | 1 | 47500867219880800 | 237504336099404000 |
| | | 2 | 0.0009784698777739363 | 5 | 1 | 47500867219880800 | 237504336099404000 |
| | | 2 | 0.0009786505252709077 | 1375 | 1 | 172730426254112 | 237504336099404000 |
| | | | ... | | | | |
| 2375043360994040013 Brooklyn optimizer >60 qubits | 1 | 3 | 0.0157480323687232 | 29 | 1 | 8189804693082897 | 237504336099404013 |
| | | 3 | 0.2581787109393368 | 2059 | 1 | 115349361874407 | 237504336099404013 |
| | | 3 | 0.0012207180270707176 | 1 | 29 | 237504336099404013 | 8189804693082897 |
| | | 3 | 0.3129888037256024 | 29 | 1 | 8189804693082897 | 237504336099404013 |
| | | 3 | 0.031251909334969276 | 1 | 29 | 237504336099404013 | 8189804693082897 |
| | | 3 | 0.5178222657414153 | 2059 | 1 | 115349361874407 | 237504336099404013 |
| | | 3 | 0.0023498535156321345 | 71 | 29 | 3345131494357803 | 8189804693082897 |
| | | 3 | 0.6562500151630974 | 29 | 1 | 8189804693082897 | 237504336099404013 |
| | | 3 | 0.0498657245188987 | 29 | 1 | 8189804693082897 | 237504336099404013 |
| | | 3 | 0.00025177709508538510 | 71 | 29 | 3345131494357803 | 8189804693082897 |
| | | 3 | 0.06483697937801486 | 29 | 1 | 8189804693082897 | 237504336099404013 |
| 2375043360994040017 Brooklyn optimizer >60 qubits | 1 | 2 | 0.015625037253130372 | 7 | 1 | 33929190871343431 | 237504336099404017 |
| | | | ... | | | | |
| | 1 | 3 | 0.2500038146977204 | 7 | 1 | 33929190871343431 | 237504336099404017 |
| | | | ... | | | | |
| 15785823750433609940401 3 Washington optimizer >80 qubits | 1 | 2 | 0.2500458658905702 | 193 | 1 | 81791832903801087774 1 | 15785823750433609940401 3 |
| | | | 0.01564037799835205 | 193 | 1 | 81791832903801087774 1 | 15785823750433609940401 3 |
| | | | ... | | | | |
| 23751578582375043360 9940401323457 Washington optimizer >100 qubits | 1 | 2 | 0.0625000298032461 | 3 | 1 | 7917192860791681120 3313467107819 | 23751578582375043360 9940401323457 |
| | | | | 261 | 1 | 91002216790708978 3946131805837 | 23751578582375043360 9940401323457 |
| | | | | ... | | | |
| | | | | 87 | 1 | 27300665037212693 51838395417511 | 23751578582375043360 9940401323457 |
| | | | | ... | | | |
| | | | | 159 | 1 | 149380997373427945 6666291832223 | 23751578582375043360 9940401323457 |
| | | | | ... | | | |
| | | | | 477 | 1 | 497936657911426 485555430610741 | 23751578582375043360 9940401323457 |
| | | | | ... | | | |
| 23751578582375043360 994040132345711 Washington optimizer >110 qubits | 1 | 3 | 0.12707519717515225 | 121 | 1 | 1962940378708681 2695036396803591 | 23751578582375043360 994040132345711 |
| | | | | ... | | | |
| | | | 0.50000028628855944 | 45594373 | 1 | 520932233948584 91333994307 | 23751578582375043360 994040132345711 |
| | | | | ... | | | |
| | | | | 376813 | 1 | 6303280030777877 451413311147 | 23751578582375043360 994040132345711 |

## 4. Concluding remarks

Considering the factorization of a large constant is a challenging problem that can be addressed using a modular multiplier based on phases to build block of a modular exponentiation circuit. Such approach is theoretically accurate taking advantages of the well-known phase definition ($\varphi_{a^{2^j}} = (2^{j-1} mod\ N).\frac{2.\pi}{N}.a$) and particularly useful.

Because of the effectiveness of our models for modular computation that defines compact circuit in both number of qubits and number of gates, our model has permit successful experimentations on physical quantum optimizer for very large integer numbers requiring circuit with more than 100 qubits.

Appendix. Examples of runs

Fig. A1. Run on Brooklyn physical quantum optimizer for 1 593 389 363

Fig. A2. Run on Brooklyn physical quantum optimizer for 237 504 336 099 404 000 (circuit of 63 qubits)

Fig. A2. Run on Brooklyn physical quantum optimizer for 237 504 336 099 404 017 (circuit of 63 qubits)